# Constitutional Law and AI Governance:

# Constraints on Model Licensing and Research Classification

Alex Mark[1] and Aaron Scher[2]

December 2025

[1] Cambridge Boston Alignment Initiative
[2] Machine Intelligence Research Institute

## Abstract

Transformative AI systems may pose unprecedented catastrophic risks, but the U.S. Constitution places significant constraints on the government's ability to govern this technology. This paper examines how the First Amendment, administrative law, and the Fourteenth Amendment shape the legal vulnerability of two regulatory proposals: model licensing and AI research classification. While the First Amendment may provide some degree of protection for model algorithms or outputs, this protection does not foreclose regulation. Policymakers must also consider administrative legal requirements, due to both agency review and authority. Finally, while substantive due process and equal protection pose minimal obstacles, procedural due process requires the government to clearly define when developers vest a legal interest in their models. Given this analysis, effective AI governance requires careful implementation to avoid these legal challenges.

## Executive Summary

This paper analyzes potential litigation risks for AI model licensing and research classification in the United States, focusing on the First Amendment, administrative law, and the Fourteenth Amendment. While policymakers have already proposed these regulatory schemes, they must consider potential legal challenges before enacting them into law.

A model licensing regime could prevent the deployment of dangerous AI models. Existing regulatory templates—including FAA type certification, FDA staged approval, and NRC accident prevention requirements—offer tested frameworks for implementing AI licensing. However, regulators will need to ensure that licensing adheres to the Constitution. To prepare for the legal implications of model licensing, policymakers should consider that the First Amendment likely protects model algorithms, but not weights. The First Amendment also likely protects outputs that humans use LLMs to create, but it will not protect LLMs as separate legal entities with distinct speech rights.

Even so, First Amendment protection for LLM development or outputs does not foreclose regulation. Content-neutral restrictions need only satisfy intermediate scrutiny, while content-based restrictions must satisfy strict scrutiny. To implement these policies, the Major Questions Doctrine and modern administrative law will require Congress to delegate regulatory authority to the appropriate executive agency. Congress should therefore consider explicit exemptions to administrative procedural statutes to allow model regulators to match the pace of technical progress. While due process concerns for model licensing are relatively limited, regulators should still clearly delineate what "entitlement" an AI lab has for model development and when that entitlement attaches.

In addition to model licensing, other policymakers have suggested classifying AI research. The Atomic Energy Act (AEA), which classifies sensitive nuclear research, provides an analogous precedent. While First Amendment protection for software or LLM development remains an open question, the First Amendment will clearly protect research publications as a form of speech. Enforcing a classification scheme could therefore require prior restraints on publication. This presents a serious constitutional issue, given the presumptive unconstitutionality of prior restraint. To overcome this burden, the government must demonstrate that publication would create an imminent threat to national security.

**Finally, this paper offers the following recommendations:**

- Assuming LLM software or outputs are protected by the First Amendment, model licensing must carry procedural safeguards, including clear standards for approval or denial, prompt decision-making, the availability of judicial review, and burden on the government to justify denial.
- The Major Questions Doctrine and Loper Bright demand that an executive agency have regulatory latitude to impose research classification or model licensing. No agency currently has this authority, so Congress must vest this authority in a new or existing agency.
- To match the pace of capabilities progress, Congress should explicitly exempt model licensing and a research classification scheme from OIRA review and notice-and-comment rulemaking.
- To avoid being struck down on procedural due process grounds, a licensing scheme should clearly delineate when an "entitlement" begins.

## Introduction

Transformative AI is coming, amplifying the potential for global catastrophic risks.[3] Thus, the U.S. government may seek proactive measures to mitigate these risks. This paper will provide a legal analysis of two proactive policies: model licensing and research classification. These policies raise novel legal issues of machine speech, expressive software, accelerated rulemaking, and technological entitlements. If the government acts without considering questions of law, its policies risk litigation or failure.

If the U.S. government seeks to constrain the greatest risks from transformative AI, reactive policies may be too late. Yet proactive policies would be exceptional, as the government has typically responded to technological change only after a catastrophe occurs. The FAA was created in response to the 1956 Grand Canyon midair collision, the deadliest aviation incident at that point.[4] More recently, Operation Warp Speed produced the COVID-19 vaccine in record time, but it began two months after the WHO had already declared a global pandemic.[5] In the wake of the Great Recession, Congress passed the Dodd-Frank Act only after the unemployment rate had peaked at 10% the previous year.[6] There are numerous other examples of societal resilience in the wake of catastrophe. Yet resilience may prove impossible against a technical system that eclipses general human intelligence.

Model licensing and research classification present two potential forms of proactive AI governance. Both approaches have been proposed to reduce catastrophic risks from misaligned superintelligent AI systems as well as the risk of catastrophic misuse by rogue actors.[7] Licensing may gate AI development through a government approval process before superintelligent AI systems have been created. Similarly, the government may seek to classify AI research methods to prevent their spread to malicious actors. These policies will also strengthen U.S. national competitiveness. Additionally, governments may see AI as a key strategic technology and act to prevent rivals from obtaining intellectual property.[8]

This paper analyzes these questions across three doctrinal domains. Part I examines First Amendment constraints, arguing that algorithms likely receive protection while weights do not, and that research classification schemes implicate the prior restraint doctrine. Part II addresses administrative law, showing that the Major Questions Doctrine and *Loper Bright* require explicit congressional delegation. Part III considers Fourteenth Amendment concerns, focusing on procedural due process requirements for licensing schemes. The paper concludes with recommendations for designing regulations that can withstand judicial review.

---

[3] Anderljung, Markus, Joslyn Barnhart, Anton Korinek, Jade Leung, Cullen O'Keefe, Jess Whittlestone, et al. "Frontier AI Regulation: Managing Emerging Risks to Public Safety." arXiv, November 7, 2023. https://arxiv.org/abs/2307.03718.

[4] Federal Aviation Administration. "Lockheed L-1049 Super Constellation and Douglas DC-7: Trans World Airlines Flight 2, N6902C; United Airlines Flight 718, N6324C, Grand Canyon, Arizona, June 30, 1956." Lessons Learned: Transport Airplane Accidents, n.d. Accessed August 7, 2025. https://www.faa.gov/lessons_learned/transport_airplane/accidents/N6902C.

[5] Ghebreyesus, Tedros Adhanom. "WHO Director-General's Opening Remarks at the Media Briefing on COVID-19—11 March 2020." World Health Organization, March 11, 2020. https://www.who.int/director-general/speeches/detail/who-director-general-s-opening-remarks-at-the-media-briefing-on-covid-19---11-march-2020; U.S. Department of Health and Human Services. "Explaining Operation Warp Speed." Fact sheet, August 2020. https://covid19.nihb.org/wp-content/uploads/2020/08/Fact-sheet-operation-warp-speed.pdf.

[6] Bureau of Labor Statistics. "The Recession of 2007–2009." BLS Spotlight on Statistics, February 2012.

[7] Barnett, Peter, Aaron Scher, and David Abecassis. "Technical Requirements for Halting Dangerous AI Activities." arXiv, July 13, 2025. https://arxiv.org/abs/2507.09801.

[8] Jensen, Benjamin. "Protecting Our Edge: Trade Secrets and the Global AI Arms Race." Congressional testimony. Center for Strategic and International Studies (CSIS), May 7, 2025. https://www.csis.org/analysis/protecting-our-edge-trade-secrets-and-global-ai-arms-race.

## Model Licensing

To prevent catastrophe, misuse, and a loss of national competitiveness, frontier AI models could be released—whether through API access, open-source release, or both—only when the government can assure the public that they are aligned and safe. While the National Emergencies Act or the International Emergency Economic Powers Act (IEEPA) may grant the power to restrict dangerous AI in a true emergency, model licensing could attempt to prevent an emergency far in advance. This policy would employ a government licensing and certification framework to legally clear a product for development, production, and marketing. Before regulators would permit companies to market their products, companies would demonstrate safety with extensive testing, threat modeling, and analysis.[9] For the purpose of this analysis, model licensing would apply to training, model weight release, and API access.

To illustrate how model licensing could apply to frontier AI, consider three examples: the Federal Aviation Administration (FAA), the Food and Drug Administration (FDA), and the Nuclear Regulatory Commission (NRC).

The FAA's type certification is the "approval and design of the aircraft and all component parts," while production certification is the "approval to manufacture duplicate products under an FAA-approved type design." Previous research shows how FAA certification could map to frontier AI regulation.[10] While the purpose of LLM use is quite distinct from civilian aviation, the FAA certification process still presents a useful scheme for gating training, model weight release, and API access.

While the FAA provides a model for certification, the FDA provides a model for gating development.[11] The FDA drug development process begins with discovery before leading to preclinical research, clinical research, agency review, and finally, post-market safety monitoring.[12] While models are not "discovered" in the same sense as drugs, this multi-stage approach would ensure that AI developers are engaged with regulators at each stage of the development process, thereby minimizing the unexpected post-approval appearance of novel risks.

Finally, the NRC model provides a useful example of accident prevention. Its requirement that nuclear reactors include post-construction accident prevention features is directly applicable to AI. To build nuclear reactors, developers must comply with regulatory requirements for both construction permitting and operating licensure.[13] The inclusion of accident prevention features in frontier AI, such as air-gapping and exfiltration prevention, will be necessary to ensure safety compliance.

---

[9] Buhl, Marie Davidsen, Gaurav Sett, Leonie Koessler, Jonas Schuett, and Markus Anderljung. "Safety Cases for Frontier AI." arXiv, October 28, 2024. https://arxiv.org/abs/2410.21572.
[10] Salvador, Cole. "Certified Safe: A Schematic for Approval Regulation of Frontier AI." arXiv, August 12, 2024. https://arxiv.org/abs/2408.06210.
[11] Lenhart, Anna, and Sarah Myers West. Lessons from the FDA for AI. New York: AI Now Institute, August 1, 2024. https://ainowinstitute.org/wp-content/uploads/2024/08/20240801-AI-Now-FDA.pdf.
[12] U.S. Food and Drug Administration. "The Drug Development Process." Last updated January 4, 2018. https://www.fda.gov/patients/learn-about-drug-and-device-approvals/drug-development-process.
[13] U.S. Nuclear Regulatory Commission. "Backgrounder on Nuclear Power Plant Licensing Process." July 2020. https://www.nrc.gov/reading-rm/doc-collections/fact-sheets/licensing-process-fs.html.

Model licensing would attempt to operationalize technical guardrails against frontier AI risks. This comprehensive screening process would follow the model of the FAA, the FDA, and the NRC to ensure that frontier models are developed safely with robust technical guardrails. The technical details of a licensing framework are beyond the scope of this paper, but any framework must comply with existing legal principles.

## AI Research Classification

In addition to risks from the production of frontier AI systems, the research behind frontier AI development may also create novel risks. It could then be treated as sensitive as research needed to produce weapons of mass destruction. While licensing would apply to the engineering of model creation, research classification would apply to the scientific discoveries that necessarily advance model creation.

In the decades leading up to World War II, the European scientific community feared its research could be used by Nazi Germany to produce an atomic weapon.[14] Reportedly led by Hungarian physicist Leo Szilard, the community took the radical step of withholding any results from publication that would have accelerated the bomb's development.[15] Despite these efforts, the foundational science necessary for the breakthrough of the atomic bomb was not secret—it had been freely discovered over the preceding decades. Atomic research required discovery, but atomic weapons engineering required infrastructure. While the government could supply this infrastructure, it also demanded—and institutionalized—research secrecy.

The advent of research secrecy led to the creation of the National Defense Information (NDI) category, which encompassed all classified information.[16] The success of restricted data and born secrecy may be indirect. Nuclear weapons have only been used in combat twice. While nine sovereign states are confirmed to possess nuclear weapons, there have been no widely publicized reports of non-state possession. The Atomic Energy Act (AEA) provides a ready precedent for securitizing sensitive AI research and containing the proliferation of dangerous AI.

---

[14] "Contrary to perhaps what is the most common belief about secrecy, secrecy was not started by generals, was not started by security officers, but was started by physicists. And the man who is mostly responsible for this certainly extremely novel idea for physicists was Szilard." [E. Fermi, Physics Today 8, 12-16 (Nov. 1955), p. 13].

[15] "The scientists did, in fact, withhold several significant papers from publication, Szilard perhaps being the first to do so, in February 1940. 3 Probably the most important of those papers to be voluntarily withheld was a report concerning the neutron-absorption cross-section of carbon. That report indicated that carbon (e.g., graphite) would be an excellent moderator for a nuclear reactor. The subsequent, very successful, U.S. nuclear-reactor effort was therefore oriented toward using graphite as a moderator. German scientists made similar measurements, obtained erroneous results, and concluded that graphite was not a good moderator.4 For the remainder of World War II, the Germans ignored graphite as a moderator, turning instead to heavy water (deuterium oxide). They were never able to obtain sufficient quantities of heavy water to do key experiments. Had German scientists learned of the U.S. results, their efforts to develop nuclear weapons might have been significantly different from their actual program." Quist, Arvin (2002). "4". Security Classification of Information: Vol. 1. Introduction, History, and Adverse Impacts, chapter 4, "Classification Under The Atomic Energy Act" (PDF). Oak Ridge National Laboratory. Archived from the original on September 22, 2021.

[16] Wellerstein, Alex. "Can Trump Just Declare Nuclear Secrets Unclassified?" Lawfare (blog), August 18, 2022. https://www.lawfaremedia.org/article/can-trump-just-declare-nuclear-secrets-unclassified.

Classifying AI research could be applied to AI development in a manner analogous to how the AEA addressed the dangers of nuclear proliferation. Even if the research is done outside of government, a classification scheme could prevent sensitive AI research from falling into the hands of rogue actors. In particular, certain government agencies would be given the authority to classify different forms of research. Publication of this research would face similar consequences to the publication of other classified information.

Classification would present a radical, yet precedented, step for research secrecy. By employing a similar framework to the Atomic Energy Act, AI research classification could rely on longstanding administrative enforcement mechanisms. Still, even existing classification schemes rest on uncertain First Amendment grounds, as this paper will discuss.

## The First Amendment

This analysis will first examine how the First Amendment applies to LLMs: first, through their development, and then through their outputs. If the First Amendment does apply to AI research and development, then specificity, both in model licensing and research classification, will determine whether these policies are vacated by the courts or remain in place long enough to constrain AI risks.

### The First Amendment Likely Protects Algorithms, but Not Weights

To approve the release of AI models, regulatory agencies may impose requirements throughout the development process. These requirements could affect model weights, training algorithms, inference algorithms, and other layers of model scaffolding. While model hardware falls outside the scope of the First Amendment, model software falls within a gray area of First Amendment protection. The following analysis examines the lengths and limitations of the First Amendment in protected AI software development.

In *Bernstein v. U.S. Dept. of State*, the plaintiff developed an encryption algorithm called "Snuffle."[17] The Arms Export Control Act authorized the President to control the import and export of defense articles and defense services by designating such items to the United States Munitions List (USML). The USML covered "cryptographic systems, equipment, assemblies, modules, integrated circuits, components, or software with the capability of maintaining secrecy or confidentiality of information or information systems."[18]

Bernstein argued that due to the AECA, he was "not free to teach the Snuffle algorithm, to disclose it at academic conferences, or to publish it in journals or online discussion groups."[19] Thus, he argued, the licensing requirements infringed on his free expression. He argued that Snuffle was more than a tool; it was communicative and a form of academic expression. The district court agreed, finding that software is a "topic of speech" communicated in cryptographic research.[20]

The Ninth Circuit made a distinction between source code—code that communicates with a compiler but is also intended for human analysis and understanding—and object code—code that communicates solely to a computer but is not interpretable by a human.[21] It ruled that source code is a form of expression and thus subject to First Amendment protections. The Ninth Circuit did not rule on whether object code is also protected in this manner.[22]

However, *Bernstein* is not binding appellate precedent. When the Ninth Circuit withdrew the opinion to hear the case *en banc*, the government mooted the appeal by changing the underlying regulations before that rehearing could occur.[23] Thus, the First Amendment implications of software remain unsettled. Since *Bernstein*, courts have analyzed these speech protections depending on the degree of functionality in the software.[24] Even taken as persuasive authority for future courts, *Bernstein*'s holding narrowly applies to a form of software directly used in academic research. "Snuffle" was not protected because software is automatically speech; "Snuffle" was protected because it was used in a speech-like manner.

[17] *Bernstein v. U.S. Dep't of State*, 945 F. Supp. 1279, 1283 (N.D. Cal. 1996).
[18] See International Traffic in Arms Regulations (ITAR), United States Munitions List, 22 C.F.R. § 121.1, Category XIII(b)(1) (Information Security).
[19] *Bernstein* at 1285.
[20] *Id.* at 1287.
[21] *Bernstein v. U.S. Dep't of Justice*, 176 F.3d 1132, 1141 (9th Cir. 1999).
[22] *Id.* at 1142.
[23] *Bernstein v. U.S. Dep't of Justice*, 176 F.3d 1132 (9th Cir. 1999), *reh'g granted, opinion withdrawn*, 192 F.3d 1308 (9th Cir. 1999), *and remanded*, No. 97-16686 (9th Cir. Jan. 26, 2000).

This functionality distinction is critical for AI. Like a skyscraper, which combines functional structures with expressive architecture, AI models contain both. AI weights comprise trillions of inscrutable parameters and are functionally equivalent to object code; they serve as machine instructions rather than human communication. Conversely, training and inference algorithms are written in legible languages like Python, debated by engineers, and taught academically. Because these algorithms function as text-based communication between humans, they likely warrant First Amendment protection similar to the source code in Bernstein, whereas the model weights do not.

An anti-regulatory view would argue that it is untenable to separate the functional and expressive components of LLM design. They might argue that an LLM's functional architecture is inseparable from the components that allow it to express outputs. Yet under the hood, LLM development is not an expressive endeavor. It is absurd to claim that trillions of parameters can be readily discussed in the same sense as "Snuffle" in *Bernstein*. The weights of the model are not the subject of parameter-by-parameter discussion. Moreover, weights are not used to express ideas between people—they are statistical representations of how the model maps inputs to outputs. They serve as instructions to a machine, not a human. In this sense, weights are more similar to the "object code," which the Ninth Circuit declined to grant First Amendment protection in *Bernstein*.

The training and inference algorithms, however, are written, analyzed, and refined by humans. These tools are readily discussed between engineers. They are techniques similar to or exactly those learned in graduate-level academic courses. Most importantly, they are text-based. Python, the language widely used for these algorithms, is as readable as any other form of text, even if it requires some translation for lay observers. Based on precedent and common sense, algorithms used in the training and development of LLMs are likely protected under the First Amendment.

## LLM Outputs May Currently Be Treated as Amplified Human Speech

The clearest argument for protecting LLM outputs is in their name: they are Large *Language* Models.[25] Neither the word "human" nor "person" is contained in the First Amendment, and the Supreme Court has previously extended freedom of speech to other non-human entities, such as corporations.[26] Courts are unlikely to grant First Amendment

---

[24] *Universal City Studios, Inc. v. Corley*, 273 F.3d 429, 454 (2d Cir. 2001) ("code can simultaneously have functional and expressive elements, with First Amendment protection applying to the expressive aspects."); *Junger v. Daley*, 209 F.3d 481, 485 (6th Cir. 2000) (holding that "[b]ecause computer source code is an expressive means for the exchange of information and ideas about computer programming . . . it is protected by the First Amendment…[but] "[t]he functional capabilities of source code, and particularly those of encryption source code, should be considered when analyzing the governmental interest in regulating the exchange of this form of speech.").

[25] See Wittes, B. (2023, March 31). A machine with First Amendment rights. Lawfare. https://www.lawfaremedia.org/article/machine-first-amendment-rights.

[26] *Citizens United v. FEC*, 558 U.S. 310, 342 (2010) ("The Court has thus rejected the argument that political speech of corporations or other associations should be treated differently under the First Amendment simply because such associations are not 'natural persons.'").

protections to LLMs as independent entities, but may instead find they amplify human speech.

First Amendment doctrine emphasizes the rights of the listener along with the rights of the speaker.[27] This concept derives from the democratic theory of the First Amendment. Alexander Meiklejohn, a speech theorist, treats spoken word as the atomic component of democracy.[28] To expand on this point, Robert Post, another speech theorist, argues that when certain forms of speech *advance* public discourse, they should receive a heightened degree of protection.[29] Robert Bork offers a narrower view, claiming that the democratic theory of the First Amendment implies that only "political speech" should be specially exempt from regulation. In his view, political speech includes "evaluation, criticism, electioneering and propaganda," but excludes "scientific, educational, commercial or literary expressions."[30]

As a general-purpose technology, LLMs could both tractably improve and confuse public discourse. Their capacity to undermine the democratic process is already evident from deepfakes, robocalls, and epistemic degradation.[31] Nor are LLMs independent members of the electorate—they are tools used by voters, not voters themselves. The potential for democratic erosion is not necessarily a basis for denying First Amendment protection (defamation may not be protected speech, but propaganda is).[32] LLMs could influence or advance one-sided issue campaigns, exacerbating current misinformation dynamics, but this ecosystem has existed for over a decade without restriction. Frontier AI carries unprecedented potential for amplified misinformation, but no existing precedent has held that misinformation justifies restricting speech.[33]

The Supreme Court ("the Court") has pointedly addressed that spending money to influence the democratic process is protected speech. In *Citizens United v. F.E.C.*, it reasoned that "the appearance of influence or access, furthermore, will not cause the electorate to lose faith in our democracy."[34] As corporations are entitled to influence the democratic process through campaign funding, LLM users may be entitled to influence the democratic process through text output, image generation, and agentic publication.[35] As individuals may contribute to political campaigns that amplify their position, individuals may use LLMs to amplify their position.[36] The role of the government, one may argue, is not to arbitrate truth, but to "[entrust] the people to judge what is true."[37]

---

[27] *Richmond Newspapers, Inc. v. Virginia*, 448 U.S. 555 (1980) (holding that closing a trial to the public was unconstitutional due to its inhibitions on the First Amendment rights of the public to listen).

[28] Alexander Meiklejohn, Free Speech and Its Relation to Self-Government (New York: Harper & Brothers, 1948), 24.

[29] Robert C. Post, "The Unfortunate Consequences of a Misguided Free Speech Principle," *Dædalus* 153, no. 3 (Summer 2024): 135-148 ("One of the very great dangers hanging over the future of free speech in the United States is the present tendency of the Supreme Court to extend to all speech the protections properly due only to public discourse, and thus to use the First Amendment to impose a libertarian, deregulatory agenda on ordinary state social and economic regulations").

[30] Bork, Robert H. (1971), "Neutral Principles and Some First Amendment Problems." 47 IND. L.J. 1.

[31] See, e.g., Ali Swenson and Will Weissert, "Fake Joe Biden Robocall Tells New Hampshire Democrats Not to Vote in Tuesday's Primary," *AP News*, January 22, 2024; *Implications of Artificial Intelligence Technologies on Protecting Consumers from Unwanted Robocalls and Robotexts*, CG Docket No. 23-362, Declaratory Ruling (Fed. Comm. Comm'n Feb. 8, 2024). For a discussion on the broader erosion of public knowledge, see Elizabeth Seger et al., *Tackling Threats to Informed Decision-Making in Democratic Societies: Promoting Epistemic Security in a Technologically-Advanced World* (London: The Alan Turing Institute, 2020).

[32] *Chaplinsky v. New Hampshire*, 315 U.S. 568, 571-72 (1942); *Lamont v. Postmaster General*, 381 U.S. 301, 305-07 (1965).

[33] *United States v. Alvarez,* 567 U.S. 709, 718 (2012) (plurality opinion) ("The Court has never endorsed the categorical rule the Government advances: that false statements receive no First Amendment protection.").

[34] *Citizens United v. Federal Election Commission*, 558 U.S. 310, 314 (2010).

[35] See also *First Nat'l Bank of Boston v. Bellotti,* 435 U.S. 765 (1978).

Still, *Citizens United* hinged on the rights of the speaker, not democracy. The majority opinion compared the First Amendment rights of a media publication with those of a corporate sponsor. Both organizations consist of individuals with independent speech rights, but *Citizens United* made clear that the First Amendment extends beyond individuals. Corporations are "associations of citizens" with individual intentions.[38] This reasoning has long-standing precedent: the extension of constitutional rights to non-individuals has applied to unions, municipalities, and even ships.[39] The Court reasoned that denying constitutional protections to entities *of humans* would deny constitutional protections to the individuals within those entities.

*Citizens United*'s implications for LLM regulation are twofold. On the one hand, its holding does find that corporate expenditures—when used to amplify speech—are protected under the First Amendment. This logic could apply to using an LLM to amplify speech, and thus supports the idea that LLM outputs are extensions of the human user's speech, which is protected. On the other hand, the Court's holding rests on the idea that corporations consist of humans. Applying the democratic theory to LLM outputs shows a mixed scenario, where the level of protection depends on the level of autonomy afforded to the model.

The autonomy theory of the First Amendment is unlikely to justify LLM protections any more than the democratic theory. According to autonomy theorists, speech is essential to individual flourishing.[40] A court need not reach the question of literal LLM autonomy to analyze LLM protections under this theory.[41] Even if a court assumes that LLMs lack intentionality, a crucial requirement for expressive activity, it would still find that a human speaker could intentionally *use* an LLM.[42]

The key difference between the intentional use of an LLM and the intentional use of other expressive tools lies in the level of user control. A paintbrush gives the artist direct control over her brush strokes. By contrast, one might argue that generating AI art removes the requisite level of human control to be considered intentional. While flag burning is a protected expressive activity, *unintentionally* dropping a flag into a fire is not. As AI increases in capabilities, users may have *less* control over the output of their chosen LLMs.

Yet the First Amendment may still protect *machine* speech via the marketplace theory of the First Amendment. The marketplace theory supports the idea that a "multitude of tongues" generates truth-seeking.[43] That does not imply that this marketplace is unlimited.

---

[36] See also *First Nat'l Bank of Boston v. Bellotti*, 435 U.S. 765 (1978).
[37] *Id.* at 355.
[38] Salib, P. N. (2024, November 5). *AI outputs are not protected speech* (rev. May 21 2024). Washington University Law Review (forthcoming). SSRN. https://ssrn.com/abstract=4687558, citing *Citizens United v. Federal Election Commission*, 558 U.S. 310, 394 (2010).
[39] Samir Chopra & Laurence F. White (2011). A Legal Theory for Autonomous Artificial Agents at 157-158. University of Michigan Press.
[40] *Whitney v. California*, 274 U.S. 357 (Brandeis, J., concurring).
[41] *See* David Atkinson, Jena D. Hwang, and Jacob Morrison, *Intentionally Unintentional: GenAI Exceptionalism and the First Amendment*, preprint, arXiv:2506.05211 (submitted June 5 2025), https://doi.org/10.48550/arXiv.2506.05211; *See also* Joel Feinberg (1992). Freedom and Fulfillment: Philosophical Essays at 52. Princeton University Press.
[42] *Texas v. Johnson*, 491 U.S. 397 (1989) (finding a Texas law that prohibited flag burning targeted expressive activities particularly because it targeted the intent of such activities").
[43] *Associated Press v. United States*, 52 F. Supp. 362, 372 (S.D.N.Y. 1943), aff'd, 326 U.S. 1 (1945).

This "multitude" only extends to speakers within the jurisdiction of the United States.[44] The Constitution may apply to non-citizens, but even non-citizens must be subject to the jurisdiction of the United States. None of the marketplace, autonomy, or democratic theories of the First Amendment would claim that the Constitution applies to those beyond the complete reach of the law.

Perhaps the most compelling argument for protecting LLM outputs relates to the rights of the listener. The Supreme Court has previously held that the First Amendment applies to the listener as well as the speaker, thus implicating the audience of LLMs as well as LLM output.[45] Eugene Volokh, another First Amendment scholar, has noted that the Supreme Court has held that First Amendment protections apply even to "speech by foreign propagandists" and the speech of "dead people" precisely because the listener benefits.[46] As a *tool*, an LLM is inherently expressive, even if it transmits unintended messages. The user intends to convey a message through the LLM, even if the LLM distorts their initial intent. The outputs could be sufficiently understood by listeners. Thus, human speech amplified through an LLM does not forfeit First Amendment protection. The output would be far too similar to other forms of expression to be excluded from free speech analysis.

As an *entity*, it remains unclear and unsettled whether an LLM will ever possess the requisite intentionality to qualify as a speaker. As models advance, courts may revisit this question. For the time being, it remains to be seen if courts will ever find that LLMs, as *entities*, have full speech rights under the First Amendment.

## AI Research Classification Will Implicate Prior Restraint

Few courts have ruled on the constitutionality of research classification. This is because, for practical reasons, the government has avoided prosecuting the cases that could be litigated. Prosecuting the illegal disclosure of restricted data may require *declassifying* classified information for evidentiary purposes, risking further proliferation of sensitive material.[47] Security officials must weigh the sensitivity of this data against the need to enforce classification regimes. If prosecutions advance to trial, the government risks disclosing the very information it intends to keep secret. Thus, most security officials resolve leaks administratively or informally.[48]

When the government does prosecute leaks, it typically relies on the Espionage Act rather than the Atomic Energy Act.[49] Take Ben-Ami Kadish's 2008 arrest. Though Kadish shared over one hundred documents with an Israeli agent, including restricted data on nuclear weapons, he was charged with espionage rather than with violating the criminal provision of the AEA.[50] Pure leak cases remain rare—at least publicly—and tend to target American citizens who leaked information to other Americans.[51]

---

[44] Salib, P. N. (2024, November 5). AI outputs are not protected speech. Washington University Law Review (forthcoming). SSRN.

[45] See *Va. State Bd. of Pharmacy v. Va. Citizens Consumer Council, Inc.,* 425 U.S. 748 (1976) (finding that the First Amendment applies to "willing speakers and willing listeners alike").

[46] Eugene Volokh, *First Amendment Limits on AI Liability*, Lawfare (Sept. 27, 2024, 8:00 AM), https://www.lawfaremedia.org/article/first-amendment-limits-on-ai-liability, *citing Lamont v. Postmaster Gen.*, 381 U.S. 301 (1965).

[47] Steven Aftergood, Punishing Leaks Through Administrative Channels, *Federation of American Scientists* (May 3, 2016) (quoting ODNI General Counsel Robert S. Litt that prosecutions for unauthorized disclosures are "often beset with complications, including … the potential for confirming or revealing even more classified information in a public trial," so agencies are told to pursue "administrative investigations and sanctions" instead).

[48] Edward C. Liu & Todd Garvey, Protecting Classified Information and the Rights of Criminal Defendants: The Classified Information Procedures Act 1 (Cong. Research Serv. Apr. 2, 2012) ("*in many cases, the executive branch may resolve this tension … by simply forgoing prosecution in order to safeguard overriding national security concerns*").

[49] 18 U.S.C. section 798.

Other than prosecution, the government might turn to prior restraints as a way to prevent unauthorized disclosure. Prior restraint attempts to prevent the publication of private speech through judicial order or administrative action.[52] These actions are **presumptively unconstitutional** except for narrow circumstances.[53] As the Court established in *Near v. Minnesota ex rel. Olson*, "it has been generally, if not universally, considered that it is the chief purpose of the [freedom of the press] to prevent previous restraints upon publication."[54] National security alone is insufficient to justify a prior restraint; there must be an imminent threat to national security.[55]

In the seminal Pentagon Papers case, *New York Times v. United States*, the government sought to restrain the New York Times from publishing classified Department of Defense research.[56] While the research was already classified, this fact did not prevent the New York Times from publishing it. Thus, it was not the classification itself that constituted a prior restraint, but a court injunction enforcing it would have. The Court did not enjoin publication in this case because of the exceptionally high burden for justifying this action.

Only one case, *United States v. The Progressive, Inc.*, squarely addressed the AEA's *born secret* provision. Born secrets encompass information so sensitive that it is automatically classified. *Progressive* offers a rare example of a court-enjoined prior restraint.[57] Before the case began, The Progressive magazine planned to publish an article supposedly containing the recipe for a hydrogen bomb. The article was titled "The H-Bomb Secret: How We Got It—Why We're Telling It." In response, the government sought a temporary restraining order against the magazine to prevent the dissemination of the bomb's recipe. Agreeing with the government and enjoining publication, the district court noted that surely the First Amendment was not so absolute as to risk the cessation of all future speech—"information of sufficient destruction potential to nullify the right to free speech and to endanger the right of life itself."[58]

*Progressive* is also not binding on courts beyond the jurisdiction of its original court, but born secrets have remained a nearly unchallenged area of national security law.[59] *Progressive* is the only case to invoke the First Amendment implications of classified nuclear research directly. The Supreme Court has never ruled on the issue, and the *Progressive* case was declared moot after the district court decision. The government ultimately dismissed the case after another newspaper, the *Madison Press Connection*, published a letter complete with the sources for the development of the bomb. It is unclear whether the Supreme Court would have upheld the district court's decision.[60] Thus, while the danger of nuclear proliferation rivals the proliferation of superintelligence, there is no clear precedent that speculative risks—even existential ones—give rise to a valid prior restraint.

[50] Katherine L. Herbig, *The Expanding Spectrum of Espionage by Americans, 1947–2015*, Technical Report 17-10 (Seaside, CA: Defense Personnel and Security Research Center, Office of People Analytics, August 2017), iii.
[51] *Id.* at 65.
[52] *Alexander v. United States*, 509 U.S. 544, 550 (1993) (defining prior restraint as "administrative and judicial orders forbidding certain communications when issued in advance of the time that such communications are to occur," with temporary restraining orders and injunctions as "classic examples of prior restraints.").
[53] *Org. for a Better Austin v. Keefe*, 402 U.S. 415, 419 (1971).
[54] *Near v. Minnesota*, 283 U.S. 697, 713 (1931).
[55] *New York Times Co. v. United States*, 403 U.S. 713, 714 (1971) (per curiam).
[56] *Id.*
[57] *United States v. Progressive, Inc.*, 467 F. Supp. 990 (W.D. Wis. 1979).
[58] *Id.* at 991.
[59] *But see* Morland, H. (2005). Born secret. Cardozo Law Review, 26(4), 1401–1408 ("Since the born secret doctrine prohibits discussion of the utilization as well as the manufacture of nuclear weapons, the only public policy that has ever risked the survival of the nation has been exempted from the First Amendment.").

## Model Licensing Would Contain Both Content-Based and Content-Neutral Restrictions

Unlike a prior restraint, model licensing is far from censorship. Licensing regimes do not strictly prevent "publication" of dangerous material. Safety tests don't seek to censor LLM speech so much as contain misaligned conduct such as power-seeking, scheming, and runaway capabilities. Still, critics may argue that conduct and speech are too similar in the LLM context. Arguably, mandatory safety tests would have the side-effect of censoring LLM speech. Even if these safety tests focus on illegal or dangerous conduct, they may change LLM behavior in what amounts to either compelled speech or censorship.

Yet if a court needs an example of true LLM censorship, DeepSeek R1 provides one.[61] When users prompt DeepSeek R1 with requests to criticize the Chinese Communist Party or discuss the Tiananmen Square protests, the model will request that the user "talk about something else." Unlike DeepSeek's outright prohibition on criticizing the CCP, preventing a model from generating bioweapon instructions is more akin to banning conduct than banning speech.[62] Admittedly, LLM training may suppress specific text outputs and may have poorly understood side effects, but the goal of safety training is to prevent catastrophic misuse through user conduct.

The distinction between safety and censorship lies in the disparate treatment of "**content-neutral**" and "**content-based**" speech regulation. Content-based laws restrict the communication of certain views or topics. For example, a statewide law banning the display of cannabis imagery in public service announcements is a content-based restriction. These laws are presumptively unconstitutional unless they clear *strict scrutiny*, meaning they are narrowly tailored to serve a compelling governmental interest.[63]

The Supreme Court has previously found national security to be a compelling interest. Consider *Holder v. Humanitarian Law Project*, in which the Humanitarian Law Project sought to provide conflict resolution training to designated terrorist groups.[64] The Supreme Court upheld this prohibition, finding that "the government's interest in combating terrorism is an urgent objective of the highest order."[65] The crucial point for the Court in *Holder* was the fungibility of support for terrorist groups. The Humanitarian Law Project's "material support" was not dangerous as an end—it provided designated terrorist groups with additional support to amplify their allegedly violent ends. Unmitigated model release would carry similar concerns. If a developer released an unlicensed open-source model with dangerous capabilities, this model would erode resource constraints for dangerous actors.

---

[60] Id.
[61] Yang, Z. (2025, January 31). Here's how DeepSeek censorship actually works—and how to get around it. *WIRED*. https://www.wired.com/story/deepseek-censorship/.
[62] Burman, T. (2025, January 27). DeepSeek AI refuses to criticize Xi Jinping: "Talk about something else." *Newsweek.*
[63] *Reed v. Town of Gilbert*, 576 U.S. 155 (2015); *City of Austin v. Reagan National Advertising*, 596 U.S. 61 (2022).
[64] *Holder v. Humanitarian Law Project*, 561 U.S. 1 (2010).
[65] *Id.* (The Court goes on to reason that prohibiting the teaching of conflict resolution training is still within the least restrictive means of combating terrorism, given substantial research on how terrorist groups can use conflict resolution tactics to delay and distract their adversaries before engaging in future conflict.).

To be narrowly tailored, a regulation cannot be overinclusive or underinclusive.[66] Courts may find that regulations meet this requirement when they target speech "most likely" to implicate the government's interest.[67] The industry provides potential examples for narrowly tailored restrictions on LLM training and text. OpenAI's Preparedness Framework[68] or Anthropic's Responsible Scaling Policy[69] present incomplete but ready templates for targeted risk cases.

By contrast, a content-neutral speech restriction does not restrict certain topics but regulates the medium of speech, typically through the "time, place, and manner" of speech. For example, rather than banning the display of cannabis imagery in public service announcements, a content-neutral law might ban televised public service announcements after 10 p.m. on local public broadcasts. These laws will be upheld if they clear *intermediate scrutiny*, meaning they are substantially related to an important governmental interest.[70] A content-neutral law is not required to use the least restrictive means, but it must reasonably allow for alternative avenues of communication.[71]

While there is no bright-line rule for satisfying an "important governmental interest," examples of important governmental interests include supporting highway safety and conserving energy.[72] Since *Bernstein*, courts have frequently upheld regulations on source code precisely because they have found such restrictions to be content-neutral.[73] Previously, the Second Circuit upheld a requirement that software developers register with the Commodity Futures Trading Commission as broker-dealers when they market stock advice algorithms to retail investors.[74] More recently, courts have upheld restrictions on algorithms that facilitate the development of "ghost guns."[75]

## Conclusion

It is possible that AI regulation entirely circumvents the First Amendment. The Supreme Court has never ruled on the issue of software as speech, nor on the issue of LLM outputs. This paper anticipates that a court would find that the First Amendment protects LLM development or outputs, but that disposition is not guaranteed. There are scenarios in which an AI regulatory agency would have far more flexibility than it would under a First Amendment analysis.

---

[66] See *Ward v. Rock Against Racism*, 491 U.S. 781, 800 (1989) (stating that a government regulation is narrowly tailored "[s]o long as the means chosen are not substantially broader than necessary to achieve the government's interest.").
[67] *Williams-Yulee v. Florida Bar*, 575 U.S. 433, 446 (2015), quoting *Burson v. Freeman*, 504 U.S. 191, 209 (1992) (emphasizing that "narrowly tailored" does not mean "perfectly tailored.").
[68] OpenAI. (2025, April 15). Preparedness framework (Version 2).
[69] Anthropic. (2025, May 14). Responsible scaling policy (Version 2.2).
[70] *City of Renton v. Playtime Theatres, Inc.*, 475 U.S. 41, 47 (1986).
[71] *Ward v. Rock Against Racism*, 491 U.S. 781, 789 (1989).
[72] See *Craig v. Boren*, 429 U.S. 190, 211 (1976); *Central Hudson Gas & Elec. Corp. v. Public Service Comm'n,* 447 U.S. 557, 568 (1980).
[73] See *Universal City Studios, Inc. v. Corley*, 273 F.3d 429, 434 (2d Cir. 2001); *United States v. Elcom Ltd.*, 203 F. Supp. 2d 1111, 1121 (N.D. Cal. 2002).
[74] *Commodity Futures Trading Comm'n v. Vartuli*, 228 F.3d 94, 111 (2d Cir. 2000).
[75] *Def. Distributed v. U.S. Dep't of State*, 121 F. Supp. 3d 680, 691-96 (W.D. Tex. 2015); *Rigby v. Jennings, No.* CV 21-1523, 2022 WL 4448220, at *10 (D. Del. Sept. 23, 2022).

Still, policymakers should not assume that courts will grant lesser protection to AI models. Rather, policymakers should assume a high level of First Amendment protection to avoid litigation after the regulatory process begins.

## ADMINISTRATIVE LAW

The past five years have revolutionized U.S. administrative law. For decades, the courts gave wide regulatory latitude to executive agencies. When reviewing the broad powers of the administrative state, one of the chief questions courts consider is whether an *agency* has exceeded its statutory authority. The recent decisions *West Virginia v. EPA* (2022) and *Loper Bright v. Raimondo* (2024) have significantly narrowed the scope of agencies' power, but even before *West Virginia*, administrative procedure was plagued by red tape. For context, the **delegation doctrine** requires Congress to provide an "intelligible" standard of scope whenever it confers rulemaking authority to an executive agency.[76] Still, Congress cannot confer fundamental legislative authority upon any agency.

Neither AI research classification nor model licensing is tenable through existing agency levers. Thus, these policies' implementation depends on whether Congress grants an executive agency new authority and, crucially, whether this agency will be constrained by the timeline of administrative procedure.

## Authority Must Flow From Congress

The Major Questions Doctrine (MQD) refers to the idea that courts will require clear and explicit authorization when Congress delegates authority to regulate questions of major political or economic significance.[77] While the Court in *West Virginia* did not expressly create a clear statement rule for the MQD, it made clear that courts must now apply this principle when interpreting congressional delegation statutes.[78]

In *West Virginia*, multiple states challenged the 2015 Clean Power Plan's (CPP) "best system of emission reduction" for operating coal power plants. Through this provision, the CPP required plants to choose between decreasing their electrical production or subsidizing solar, wind, or natural gas electricity generation. The litigants did not dispute that the Clean Air Act granted the EPA authority to regulate a power plant's standard of performance or "best system of emission reduction."[79] They disputed whether the plain reading of "best system of emission reduction" included systems that reduced carbon emissions. The Court reasoned that "generation shifting"—shifting between different sources of energy—was not within the plain reading of the Clean Air Act, and thus the EPA did not have the authority to enforce generation shifting.[80]

The Court's decision hinged on the CPP's potential to "substantially restructure the energy market," a major question of political or economic importance.[81] The CPP was so consequential for the nation's economic and political system, the Court reasoned, that the EPA had to point to "clear congressional authorization" for a generation-shifting scheme.[82] The EPA could not do so. For the first time, the Court relied on the MQD as the primary basis for vacating the rules of an executive agency.

---

[76] *Whitman v. American Trucking Ass'ns, Inc.*, 531 U.S. 457, 472 (2001).
[77] *West Virginia v. Environmental Protection Agency*, 597 U.S. 4 (2022); *DA v. Brown & Williamson Tobacco Corp.* (2000) (apply "common sense" when assessing Congress's intention to delegate powers to regulatory agencies); *Utility Air Regulatory Group v. EPA* (2014); *King v. Burwell* (2015); *See also Biden v. Nebraska, 600 U.S. 477 (2023)* (relying on the MQD in part to hold that the HEROES Act of 2003 did not authorize the Department of Education to implement broad loan forgiveness programs.).
[78] *See* J. W. Hampton, Jr. & Co. v. United States, 276 U.S. 394, 409 (1928) (upholding Congress's delegation of tariff-setting authority to the President and explaining that such delegations are constitutional so long as Congress supplies an "intelligible principle" to guide the exercise of the delegated power).
[79] Clean Air Act, 42 U.S.C. §§ 7401–7671q.
[80] *West Virginia v. Environmental Protection Agency,* 597 U.S. 4 (2022)
[81] *Id.*

Shortly after *West Virginia*, the Court overturned the *Chevron* doctrine in *Loper Bright Enterprises v. Raimondo*.[83] The *Chevron* doctrine stated that if Congress is silent on a given regulatory issue, the court must only determine whether the agency's action is "based on a permissible construction of the statute," not the court's statutory construction.[84] Rather than allowing for courts to defer to agencies' statutory construction, the Court in *Loper Bright* held the Administrative Procedure Act (APA) requires courts to exercise their *independent judgment* in deciding whether an agency has acted within its statutory authority—they no longer defer to agencies' judgment in interpreting rules.[85]

Courts may still consider an agency's construction of its statutory authority, but they will no longer defer to it. *Skidmore v. Swift & Co.* provides factors the Court will consider in these cases. In *Skidmore*, the Court determined that when evaluating the validity of an agency's rulemaking authority, it should consider 1) the thoroughness in the agency's consideration of the statute, 2) the "validity of its reasoning," 3) whether the interpretation has been consistent over time, and 4) any other persuasive factors.[86]

It is almost certain that AI model licensing will implicate the MQD. The regulation of artificial intelligence is a profound question of political and economic significance. Private investment in AI surpassed $100 billion in 2024 alone.[87] Up to this point, the only federal comprehensive AI capabilities regulation has been the (now revoked) Executive Order 14110. E.O. 14110 instructed the National Institute of Standards and Technology (NIST) to devise national standards for AI transparency and best practices. NIST is statutorily a *non*regulatory agency. NIST was established by the NIST Organic Act.[88] This statute repeatedly grants NIST broad authority to set technology "standards" and "voluntary" commitments by industry professionals. However, it does not grant authority to NIST to enforce these standards or assign penalties to deviations.

The authority to license AI models does not currently exist in the U.S. administrative state. The DPA is an instrument for *increasing* critical infrastructure supply, not instituting safety regulations. Principles of statutory interpretation require that any DPA invocation aligns with legislative intent.[89] If the government wants to restrict publication of open weights, restrict API access, or classify sensitive AI research, Congress must clearly delegate that authority to an agency. While the MQD does not require that any congressional grant of authority spell out those *specific* model licensing criteria, it must be clear from the statute that the agency has authority to regulate artificial intelligence.

---

[82] *Id.*
[83] *Loper Bright Enterprises v. Raimondo*, 603 U.S. ___ (2024)
[84] *Chevron U.S.A., Inc. v. NRDC*, 467 U.S. 837 (1984)
[85] *Loper Bright Enterprises v. Raimondo*, 603 U.S. ___ (2024)
[86] *Skidmore v. Swift & Co.*, 323 U.S. 134 (1944)
[87] Stanford Institute for Human-Centered Artificial Intelligence (Stanford HAI), *The 2025 AI Index Report* (2025), https://hai.stanford.edu/ai-index/2025-ai-index-report (last visited July 20, 2025); Alex Singla et al., *The state of AI: How organizations are rewiring to capture value*, McKinsey & Company (Mar. 12, 2025), https://www.mckinsey.com/capabilities/quantumblack/our-insights/the-state-of-ai (last visited July 20, 2025).
[88] 15 U.S.C. §272
[89] *Foster v. United States*, 303 U.S. 118, 120 (1938) ("Courts should construe laws in harmony with the legislative intent and seek to carry our legislative purpose").

## A Flexible Agency Must Be Fast

A new AI regulatory agency would face a fundamental conflict between the rapid pace of technological progress and the slow nature of federal rulemaking. To withstand legal scrutiny and avoid being deemed "arbitrary and capricious" under standards like *Motor Vehicle Manufacturers Ass'n v. State Farm*, the agency must base its licensing criteria on substantial research and clear evidence. This requirement creates a bottleneck, as the deep investigation needed to justify regulations inevitably moves more slowly than the algorithmic efficiency and model improvements it seeks to govern.

Consequently, the agency runs the risk that its rules will become obsolete before they are even implemented, as measures of capability and safety shift within months. Because the administrative process naturally lags behind innovation, an AI agency must retain the power to frequently restructure its certification regimes. Without the flexibility to undergo constant review and adaptation, the agency's rigid adherence to legal standards will leave it unable to keep up with the realities of AI advancement.

Before an AI regulatory agency issues new rules, the Administrative Procedure Act (APA) requires the agency to publish the rule for notice-and-comment if the action is categorized as a substantive rule of general applicability, but not if it is an interpretive rule or a statement of general policy.[90] AI model certification and restriction criteria would almost certainly be considered substantive rules of general applicability because they concern rights and obligations and apply across the full coverage of the regulated domain. Once the agency determines that publication is required, it must draft the proposed rule. Under the APA, this drafting triggers an extensive public comment period. However, the APA exempts specific domains from this requirement, including rules concerning military or foreign affairs functions; agency management or personnel; public property, loans, grants, benefits, or contracts.

It is foreseeable that AI model licensing and research classification would implicate military or foreign affairs functions. If so, the agency could legitimately refrain from establishing a public comment period, significantly expediting the regulatory process. Yet even if exempt from notice-and-comment, Executive Order 12866 mandates that significant regulatory actions must be submitted to the Office of Information and Regulatory Affairs (OIRA) within the Office of Management and Budget (OMB) for review.[91] Significant actions include those likely to have an annual economic effect of $200 million or more, create serious inconsistencies with other agencies, or raise novel legal or policy issues. AI licensing policies will almost certainly be deemed significant enough to merit this review.

Unless Congress intervenes, this process creates a severe bottleneck. In a statutory grant of authority, Congress can exempt certain AI agency actions from OIRA review. Without such legislative carveouts, the OIRA timeline will likely outlast the relevance of the model being regulated.

Consider the standard timeline: The agency must submit its proposed rule to the OMB, with review taking up to 90 days. Subsequently, the agency must publish the rule in the Federal Register, allowing 30 to 60 days for public comment, or up to 180 days for complex rules. After the comment period, the agency must respond to feedback and draft its final rule.[92] This final rule is then subject to another OMB review, which can take an additional 90 days.[93] The final rule is only published after this exhaustive cycle of drafting, initial review, publication, comment, redrafting, and final review.

---

[90] 5 U.S.C. § 553

[91] Exec. Order No. 12,866, 58 Fed. Reg. 51,735 (Oct. 4, 1993)

There is one potential bypass: the "good cause" exception.[94] The APA allows agencies to skip notice-and-comment when they find that such procedures are "impracticable, unnecessary, or contrary to the public interest." Good cause may apply to emergencies, situations where prior notice would impede the regulation's effectiveness, or instances where Congress intends an exemption.[95] Courts have upheld good cause exceptions even years after the motivating emergency. For example, the D.C. Circuit upheld the FAA's revocation of pilot certifications for security risks, a policy crafted in response to September 11 but not implemented until 2003, because notice-and-comment would have caused unacceptable delays in addressing security threats.[96] Similarly, courts have found good cause when advance notice could distort markets, such as with the Federal Energy Administration's price-equalizing measures.[97]

## Conclusion

The Major Questions Doctrine and *Loper Bright* have changed the nature of U.S. administrative law. Judicial scrutiny is tighter than it has been in decades, and the scope and implementation of AI regulations must survive litigation. The stronger an agency's source of authority, the more likely a court will uphold the agency's rules. Congress must ensure that administrative procedure does not hinder agile AI regulation. To protect the public from the risks of frontier AI, Congress could exempt AI regulators from the burdens of administrative procedure.

---

[92] Justia, *The Notice and Comment Process Legally Provided for Agency Rulemaking* (last reviewed May 2025), https://www.justia.com/administrative-law/rulemaking-writing-agency-regulations/notice-and-comment/ (“Agencies must … respond in some form to all comments received.”).

[93] Exec. Order No. 12,866 § 6(a)(3)(D), 58 Fed. Reg. 51,735, 51,741 (Oct. 4, 1993) (requiring that “in emergency situations or when an agency is obligated by law to act more quickly than normal review procedures allow, the agency shall notify OIRA as soon as possible” and must still schedule rulemaking so as to give OIRA sufficient time for review).

[94] 5 U.S.C. § 553(b)(4)(B)

[95] Jared Cole, Cong. Rsch. Serv., The Good Cause Exception to Notice and Comment Rulemaking: Judicial Review of Agency Action 4-5 (Jan. 29, 2016).

[96] *Jifry v. F.A.A.*, 370 F.3d 1174, 1179-80 (D.C. Cir. 2004).

[97] *Mobil Oil Corp. v. Dep't of Energy*, 728 F.2d 1477, 1492 (Temp. Emerg. Ct. App. 1975); See also *Nader v. Sawhill*, 514 F.2d 1064, 1068 (Temp. Emerg. Ct. App. 1975); *DeRieux v. Five Smiths, Inc.*, 499 F.2d 1321, 1332 (Temp. Emerg. Ct. App. 1975).

## Due Process and Equal Protection

The existing means for enforcing research classification, whether through administrative, civil, or criminal means, are already extensive. If the government follows the model of the AEA, it will not need to fundamentally rethink the procedural mechanisms for enforcing AI research classifications. However, model licensing may require novel procedural approaches. Policymakers should clarify exactly when the approval process attaches in development and what process is due for model licensing.

Due process is the constitutional bedrock of *fairness*. *Procedural* due process refers to the standard of adequate protection the government must provide before depriving one of a protected interest—life, liberty, or property—through both notice and an opportunity to be heard. *Substantive* due process refers to the fundamental rights for which deprivation is subject to heightened scrutiny, even with adequate procedural protections.

### AI Regulation Will Not Implicate Substantive Due Process

While *enumerated rights*—such as speech—are directly found in the text of the Constitution, *unenumerated* rights have been interpreted through centuries of case law. There is no exact test for delineating an unenumerated fundamental right, but the Supreme Court has previously held that fundamental rights include the right to marry,[98] the right to interstate travel,[99] and the right to have custody of one's children.[100] Any restrictions on fundamental rights are subject to strict scrutiny. By contrast, economic rights have not been considered fundamental rights since the Lochner era, and therefore, restrictions on the "right to contract" are only subject to the rational basis test.[101] To clear the rational basis test, there must be a *legitimate* government interest that is *rationally related* to the regulation.[102] For a court to find the government's interest "legitimate," it need not consider the actual reason the law was enacted and need only find a conceivable justification.[103]

Fundamental rights are the exception, not the rule. While AI restrictions may implicate the First Amendment, they will not implicate any unenumerated fundamental rights. In *Washington v. Glucksberg*, the Court held that a fundamental right may exist when it is "deeply rooted in the Nation's history and tradition."[104] The *Glucksberg* decision has since received inconsistent treatment by the Court—*Obergefell v. Hodges* found that the "history and traditions" requirement was not dispositive, but *Dobbs v. Jackson Women's Health Organization* directly relied on *Glucksberg* to reach its decision. Thus, it remains to be seen how pertinent the *Glucksberg* formula is in determining the existence of a fundamental right. If one does consider the "history and traditions" of the United States, there is certainly no *right* to technological innovation. One may have an *interest* in innovation, but that interest may be subject to restrictions.

---

[98] See *Loving v. Virginia*, 388 U.S. 1 (1967).
[99] See *Paul v. Virginia*, 75 U.S. (8 Wall.) 168 (1869); see also *Saenz v. Roe*, 526 U.S. 489, 502–03 (1999).
[100] See *Santosky v. Kramer,* 455 U.S. 745 (1982).
[101] *See West Coast Hotel Co. v. Parrish*, 300 U.S. 379 (1937) (holding that "liberty of contract" is not a fundamental right).
[102] *FCC v. Beach Communications, Inc.*, 508 U.S. 307, 313 (1993).
[103] *Id.* at 315.
[104] *Washington v. Glucksberg*, 521 U.S. 702 (1997).

## Litigating Model Licensing Is Unlikely to Implicate Equal Protection

While the equal protection clause of the Fourteenth Amendment, as written, only applies to the states, it has also been reverse-incorporated[105] into the Fifth Amendment.[106] In nearly all equal protection cases, disparate treatment of an individual or group by the government will not implicate the equal protection clause unless the target of the action is a member of a "suspect classification" (e.g., racial[107] or gender[108] classifications). However, in *Village of Willowbrook v. Olech*, the Supreme Court carved out an exception for a "class-of-one."[109] When an individual can prove she was "intentionally treated differently from others similarly situated and that there is no rational basis for the difference in treatment," the government has violated her right to equal protection under the law.

In *Olech*, the Village of Willowbrook (Village) only agreed to connect the Village's water supply to Olech's property if she agreed to grant the Village a "33-foot easement." While ordinarily, Olech would not have an equal protection claim in this situation, the Village only required a 15-foot easement for other similarly situated property owners. Olech claimed the requirement of a 33-foot easement was "irrational and arbitrary" and "motivated by ill will." The Court agreed. While it declined to hold that the plaintiff must show "ill will" on the part of the government, the Court found that a plaintiff may prove an equal protection violation under the "class-of-one" theory when the basis for disparate treatment is "irrational."[110]

It is highly unlikely that a court will invalidate an AI license refusal or termination on equal protection grounds, even when invoking the "class-of-one" theory. Still, regulators must ensure that any actions adverse to AI developers conform to clear and identifiable standards. It is foreseeable that certain AI developers may take a more flippant attitude toward safety testing, rather than an industry-wide pattern of safety neglect.[111] If regulators end up denying licenses to specific developers more frequently, they should prepare for class-of-one challenges by documenting the evidence of noncompliance to prove that disparate treatment of one party is rational and not arbitrary.

## Model Licensing Will Implicate Procedural Due Process

While substantive due process demands fairness in the type of interest the government deprives, procedural due process demands fairness from the system of deprivation. The Court has determined that the government can meet due process requirements with notice,[112] a hearing,[113] and an impartial decision maker.[114] Whenever the government deprives an actor of its private interest, a court will balance the importance of the private interest affected, the risk of erroneous deprivation, the value of any additional or substitute procedural safeguards, and the importance of the state interest involved.[115]

---

[105] *Bolling v. Sharpe*, 347 U.S. 497 (1954).

[106] The doctrine of incorporation holds that the Fourteenth Amendment's Due Process Clause makes most of the Bill of Rights applicable to the states. In contrast, reverse incorporation applies rights to the federal government, even though the Constitution expressly directs them at the states.

[107] *Bolling v. Sharpe,* 347 U.S. 497 (1954).

[108] *Reed v. Reed*, 404 U.S. 71 (1971).

[109] *Village of Willowbrook v. Olech*, 528 U.S. 562 (2000).

[110] *Id.* at 565; *See also Sioux City Bridge Co. v. Dakota County,* 260 U.S. 441, 447 (Holding that to prove discrimination, the plaintiff must show more than "mere errors of judgment" and instead show "an intentional violation of the… principle of practical uniformity.").

[111] Kyle Wiggers, *xAI's Promised Safety Report Is MIA*, TechCrunch (May 13, 2025, 3:02 PM PDT), https://techcrunch.com/2025/05/13/xais-promised-safety-report-is-mia/; Maxwell Zeff, *OpenAI Ships GPT-4.1 Without a Safety Report*, TechCrunch (Apr. 15, 2025, 9:12 AM PDT), https://techcrunch.com/2025/04/15/openai-ships-gpt-4-1-without-a-safety-report/.

[112] See *Mullane v. Central Hanover Bank & Trust Co.*, 339 U.S. 306, 314 (1950) ("an elementary and fundamental requirement of due process in any proceeding which is to be accorded finality is notice reasonably

Any licensing scheme must consider the appropriate safeguards for preventing unfair deprivation of an AI company's private interest. The private interest, in the case of AI restriction, will likely derive from the commercial value of restricted models. Alternatively, if AI research is classified, academic researchers or institutions would hold a private interest. Their interest would derive from their freedom to seek information, which the First Amendment may implicate. However, AI research classification and declassification could mirror the procedural requirements of declassification in the AEA of 1954.

For this section, consider the state's interest narrowed to that of protecting the public from misaligned frontier models. Courts are unlikely to find that AI developers have a "private interest" in model development *before* these models exist, but policymakers should still anticipate litigation. Even if a developer has not deployed the model—or even trained it—the costs sunk into development before training could create the expectation of an entitlement. Even before training, the scaffolding, algorithmic design, and model architecture exist. **Regulation must clearly indicate the stage at which an entitlement begins.**

Licensing authorities would be wise to proceed with mandatory certification standards as early in the development process as model ideation.[116] Approval mechanisms may also depend on development components, such as copyright in code and datasets, trade secrets in weights, and potential patents in architecture. Thus, finding approval mechanisms that respect procedural due process depends on when the entitlement to development attaches.

The Court's holding in *Board of Regents v. Roth* makes clear that for an individual to claim a violation of procedural due process, he must prove he has a legitimate "entitlement."[117] In *Roth*, Wisconsin State University–Oshkosh informed David Roth, a one-year contract professor, that the university would not renew his contract term.[118] He was not given a justification for the lapse in his contract. The Court reasoned that, since Roth's contract ended after one year, he no longer had an "interest" in employment at the end of his contract. Thus, the due process clause did not require the state to provide him with a hearing or justification for termination. Thus, the Court held that, for a plaintiff to validly claim a property interest, he must have "more than a unilateral expectation" and instead have "a legitimate claim of entitlement."[119] Here, Roth's property interest was defined by and within the terms of his contract. Anything beyond that was outside the scope of his property interest.

Thus, any statutory scheme that restricts API access or open model release must clearly define the expectation an AI developer should have *before, during, and after* a license is certified. When designing restrictions, policymakers should consider the procedural calculated, under all the circumstances, to apprise interested parties of the pendency of the action and afford them an opportunity to present their obligations.”).

[113] Baldwin v. Hale, 68 U.S. (1 Wall.) 223, 233 (1863) (“Parties whose rights are to be affected are entitled to be heard; and in order that they may enjoy that right they must first be notified.”).

[114] *Goldberg v. Kelly*, 397 U.S. 254, 271 (1970).

[115] *Mathews v. Eldridge,* 424 U.S. 319, 335 (1976).

[116] Cole Salvador, Certified Safe: A Schematic for Approval Regulation of Frontier AI, Page 17 (Aug. 2024).

[117] *Board of Regents of State Colleges v. Roth*, 408 U.S. 564, 577 (1972).

[118] *Id.* at 564.

[119] *Id.*

challenges of allowing for indefinite licensing—this would create an indefinite entitlement. If, instead, licenses were to lapse without timely renewal, the government would avoid the burden of an indefinite entitlement.

There is already significant precedent on where licensing schemes have both conformed with and violated the due process clause. The landmark case on the erroneous deprivation of property through a licensing scheme is *Bell v. Burson*.[120] In *Bell*, the Court considered whether Georgia's Motor Vehicle Safety Responsibility Act (MVSRA) violated the plaintiff's procedural due process rights. The MVSRA provided mandatory driver's license and registration suspension after any uninsured motorist was involved in an accident, unless the uninsured motorist "posts security" to cover potential liability. Under this scheme, payment of security *would not* be considered an admission of liability, *but* would occur before any procedural hearing for determining said liability.

The Court reasoned that, since a motorist certainly had a private interest in a driver's license, Georgia must provide adequate procedural safeguards. While Georgia argued that the state has an interest in "protecting a claimant from the possibility of an unrecoverable judgment," the Court did not agree that this justified removing procedural safeguards for the motorist. Importantly, Georgia claimed that liability was not pertinent to the MVSRA's statutory scheme, since payment to the aggrieved party would not be considered an admission of liability. The Court notes that the MVSRA is still "developed around liability-related concepts." It is not a "no-fault scheme." The Court thus holds that the MVSRA violates the due process clause and that Georgia must provide an adequate forum for a determination of "reasonable possibility" of liability before terminating the motorist's license.

The MVSRA is pertinent to model licensing. If the government intends to terminate a license for an AI lab or AI model, a court may require the government to grant an AI company a hearing to determine its level of fault *before* it loses its license.[121] Unlike *Bell*, a court would not require a hearing due to the implications of liability. It is unclear how *Bell* would apply to licensure consequences of non-liability factors such as software engineering capabilities. A post-termination hearing is insufficient unless, under *Matthews*, the government's interest in license termination is so great that additional procedural safeguards would excessively burden the government.[122] If a regulator determines that a previously licensed company's model poses imminent danger to the public, there may not be enough time to give the company a pre-termination hearing. Therefore, while any regulation scheme must provide adequate safeguards for erroneous license termination, policymakers need not constrain the government to the point of eroding safeguards for public safety.

In fact, in cases of imminent danger to the public, the Court has held the state can infringe upon private interests without notice or a full opportunity to be heard. Examples of this include the seizure of diseased food,[123] tax collection,[124] and the wartime seizure of adversaries' property.[125] Evidently, there is precedent that a court will prioritize public safety over the private interest involved. Since the danger of misaligned, uncontrollable AI far exceeds the danger from diseased food and untimely tax collection, it seems likely that in exigent circumstances, a court will permit the government to terminate a model license before the AI developer has the full opportunity to be heard.

[120] *Bell v. Burson*, 402 U.S. 535 (1971).
[121] *Goldberg v. Kelly*, 397 U.S. 254, 260–63 (1970); *Cleveland Bd. of Educ. v. Loudermill*, 470 U.S. 532, 542–46 (1985).
[122] Note that the Court has even permitted states to suspend motorists' licenses without a hearing in *Mackey v. Montrym*, 443 U.S. 1, 17-18 (1979).
[123] *N. Am. Cold Storage Co. v. City of Chicago*, 211 U.S. 306 (1908).
[124] *Phillips v. Commissioner*, 283 U.S. 589, 597 (1931).
[125] *Cent. Union Tr. Co. v. Garvan*, 254 U.S. 554, 566 (1921).

## Conclusion

While model licensing is already rife with process, regulators must clarify when the entitlement to a private interest in AI development attaches. Given the multilayered AI development stream, the various stages of approval must be precise. Moreover, in cases where the government terminates a model's license, it should prepare to argue the manifest necessity for loosened procedural safeguards. A court may be willing to allow pre-notice termination, but only with evidence of an impending emergency. It is important to note that investment-backed expenditures sunk into model development may still require the government to provide just compensation for restricting a model's release under the Takings Clause of the Fifth Amendment. A Takings Clause analysis is beyond the scope of this paper.

## Conclusion

The Constitution does not prohibit proactive AI governance, but it dictates the structural limits of policy. This paper has identified the legal boundaries within which model licensing and research classification must operate to survive judicial scrutiny. These policies are not the only means of constraining dangerous AI, but both policies have precedent in other technological contexts. The First Amendment, administrative law, and the Fourteenth Amendment all set limits on the scope of AI regulation. These limits do not preclude the possibility of effective measures to contain misaligned AI, but policymakers must consider them when drafting legislation.

**This paper offers the following recommendations:**

• Assuming LLM software or outputs are protected by the First Amendment, model licensing must carry procedural safeguards, including clear standards for approval or denial, prompt decision-making, the availability of judicial review, and burden on the government to justify denial.

• The Major Questions Doctrine and *Loper Bright* demand that an executive agency have regulatory latitude to impose research classification or model licensing. No agency currently has this authority, so Congress must vest this authority in a new or existing agency.

• To match the pace of capabilities progress, Congress should explicitly exempt model licensing and a research classification scheme from OIRA review and notice-and-comment rulemaking.

• To avoid being struck down on procedural due process grounds, a licensing scheme should clearly delineate when an "entitlement" begins.

Transformative AI models may pose novel and severe public safety risks. Our regulatory system is not prepared, and the central tension is temporal. To ensure proactive regulations are effective, policymakers must consider the legal implications of these policies throughout the legislative process. Otherwise, these policies risk being vacated by the courts or defeated during litigation.

## Author Contribution Statement

Alex Mark is a Research Fellow with the Cambridge Boston Alignment Initiative. He identified the policy mechanisms, conducted the legal analysis, and was responsible for most of the writing in the project. Please direct correspondence to mark2023@lawnet.ucla.edu.

Aaron Scher was the Research Mentor for this project. He is a member of the Technical Governance Team at the Machine Intelligence Research Institute. Aaron developed the research direction for the project and provided feedback throughout the project.

## Acknowledgements

For helpful comments and discussions, the authors thank Emre Yavuz, Chris Ackerman, and Mackenzie Arnold. Any remaining views and errors are the authors' alone. This project was financially supported by the Cambridge Boston Alignment Initiative.